\documentclass[preprint,showpacs,showkeys]{revtex4}
\begin{document}

\title{The informational nature of quantum mechanics: A novel look at the interference experiment}

\author{Mohammad Mehrafarin}
\affiliation{Physics Department, Amirkabir University of Technology, Tehran 15914,
Iran; mehrafar@cic.aut.ac.ir}

\begin{abstract}
It is argued that the nature of probability is essentially informational rather than physical and that quantum mechanical predictions should be viewed as logical inferences made on the basis of the information content of a given experimental situation. By implementing such a viewpoint, it is possible to maintain a sharp distinction between the physical and statistical aspects of quantum mechanics. The idea is applied to the double-beam interference experiment, reproducing the results of the standard formulation of quantum mechanics in a manner that renders the notion of wave-particle duality superfluous.
\end{abstract}

\pacs{03.65.Ta, 03.67.-a, 02.50.Cw, 89.70.+c}
\keywords{Information, Inference, Probability, Interference experiment}

\maketitle
\section{Introduction}

In his classic EPR argument \cite{Einstein}, Einstein concluded that the quantum state of a system can not be an objective physical property of the system. Indeed, it has also an anthropomorphic nature because it is a property of the `experimental setup', i.e., the situation that we create by the experiments we choose to perform on the system. A given physical system (e.g. an electron) corresponds to many different setups 
depending on which parameters we choose to observe and/or control. In this respect, the quantum state is like the entropy in thermodynamics which is not a real property of the physical system but rather a property of the `thermodynamic system'; a given physical system corresponds to many different thermodynamic systems depending on which macrovariables one chooses to observe and/or control \cite{Jaynes}. The analogy is interesting because a quantum state, like entropy, contains information which, therefore, pertains not to the physical system but to the particular experiments we choose to perform on it. Nevertheless, quantum mechanics denys the possibility for any observer to influence the outcome of a measurement once the experimental setup is prepared and particular qualities of an individual observer do not enter the description. The setup forms the objective context and quantum mechanics predicts what unrolls from that in an observer independent manner. The distribution of the outcome is, thus, a characteristic property of the setup so the latter must be taken into account as conditioning the outcome of the experiment.

Starting from this premises, the nature of the probabilities that enter the theory can be explained in terms of the above conditioning. In its most general form, probability can be viewed as expressing the logical relation between a proposition and its conditioning information (see \cite{Jaynes2, Ballentine} and the references therein); it quantifies what is logically inferable from that information with regard to the particular proposition. The translation of a given information into a definite probability assignment is, thus, an act of inference based on the optimal processing of the available information in a manner that does not arbitrarily ignore any part of it while ensuring that no other arbitrary assumptions have been introduced. In the present context, the conditioning information is the information content of the setup itself; it consists of the preparation information together with whatever physical laws and thought arguments that are relevant to the situation under consideration. The propositions, on the other hand, are the various final results at each repetition of the experiment. The logical relation between the two reduces to a deduction when the information implies the proposition or its denial (in the limit of probability values $1$ and $0$, respectively); classical mechanics being the ultimate example of such situations. Quantum mechanically, however, the information content of a setup is, in general, inherently insufficient to yield a deduction. Objectively speaking, the degree of control in reproducing a specific result is inherently limited by the setup itself. It is precisely this incomplete information (degree of control) that makes quantum mechanical predictions (experimental results) essentially statistical (irreproducible); the unavailable information (degree of control) is of {\it no consequence} to our predictions (the actual results) other than making them statistical (irreproducible). Indeed, by making statistical predictions it is automatically implied that the missing information/degree of control, although relevant to the result of a single repetition, is irrelevant to the outcome as a distribution. Instead, it is the available incomplete information that is represented by a probability distribution and from that it is just not possible to deduce the physical mechanisms that produce a single result. Phrased objectively, a frequency distribution merely represents the limited control of the setup in reproducing a specific result and not the physical mechanisms that give rise to such a result. In other words, it is the information content/degree of control of the situation that is reflected in the actual distribution of the outcome, in the same way that physical causations are reflected in the results of reproducible trials. In short, probability distributions basically describe information and do not correspond to physical causal influences. This is in line with Bell inequality arguments to the effect that quantum mechanical probabilities can not arise out of a causal (local) theory. Adopting such a stand has profound effects on the way we interpret the predictions of quantum mechanics and the actual outcomes of irreproducible trials in general.

What we are essentially implying is that quantum mechanical predictions should be viewed as logical inferences made on the basis of the information content of a given experimental situation. Phrased objectively, they are the only outcomes compatible with the degree of control exerted by the setup itself in reproducing a specific result. In dealing with such situations, the proper question to ask is: what can be logically infered from the prior information about a setup with regard to obtaining the various final results? The (subjective) probability distribution thus assigned is expected to be observed experimentally as the frequency distribution of the outcome, if and only if the information and its logical treatment have been proper. By implementing such a viewpoint in an appropriate framework, it is possible to maintain a sharp distinction between the physical and statistical aspects of quantum mechanics. The former constitutes the information content of a setup while the latter enters the representation of this information by a probability distribution for the outcome through logical inference. This, we suggest, is how quantum mechanics should be viewed and ultimately formulated, i.e., as the theory of optimal processing of (prior) information about given experimental situations. Indeed, the very fact that the physical and statistical aspects of the theory are scrambled up in its present formulation, together with a `physical' instead of an `informational' look at the probability itself, has been the source of many paradoxes as well as the need for interpretation. In the following example, we bring the idea into sharp focus by applying it to the double-beam interference experiment (Mach-Zehnder setup) which arguably lies at the heart of quantum mechanics. Our results will coincide with those obtained via the standard formulation of quantum mechanics in a manner that does not bear the mechanism of wave-particle duality at all. 

\section{The double-beam interference experiment}

In each repetition of the experiment, a single particle is incident on a beam splitter which sends the particle along one of the two arms of length $r_1$ and $r_2$. The particle is then collected by means of two detectors, one facing each arm. Let us refer to this as setup 1. We also consider a different arrangement (setup 2), in which the two secondary beams are mixed via a second beam splitter in the standard manner. 

In each repetition of the experiment we find that only one detector clicks which means that we are detecting a single particle. This is true for both setups. However, the distribution of these clicks is a characteristic property of the setup (and not the particle) which describes its information content. Let $\{P_i\}$ denote the probability distribution of the outcome, where $i=1,2$ lables the final result (detector click) at each repetition of the experiment. We are interested in infering $\{P_i\}$ for each setup on the basis of the information that went into its preparation together with whatever physical arguments that are relevant to the situation. Let us, therefore, consider each setup separately.

Setup 1: This setup is prepared in such a manner that one can associate with every click a path (and hence a path length). This constitutes the prior information about this setup. The length of the arms could have been left unspecified as far as this setup is concerned, because if we change them arbitrarily, the prior information will not be affected in any way. Consistency, then, requires that the subjective probability assigned to this information be independent of the length of the arms. Because of the symmetry of the situation, one infers that $P_1=P_2=\frac{1}{2}$ for all $r_1, r_2$. This is the distribution that one expects to observe experimentally. It characterizes this setup in that it describes its information content in a unique manner.

Setup 2: Here the preparation is such that, because of the mixing produced by the second beam splitter, one cannot associate with every click a path nor a path length; unless, of course, $r_1=r_2$, where a path length (but not a path) can be associated. This last point, although seemingly trivial, is a part of the prior information about the setup which must not be disregarded unconsciously. It immediately implies that the prior information cannot be indifferent to the path difference $x=r_1-r_2$ because it changes when $x=0$. Consistency, then, requires that the subjective probability describing this information must depend on $x$.
We, therefore, write $P_1=f(x), P_2=1-f(x)$ where $f(x)$ is an arbitrary (non-constant) function with range $[0,1]$ such that $f(0)=0$ or $1$. (This is because the beam initially has to enter the first beam splitter along one arm or the other, and this choice for $x=0$ breaks the symmetry of the situation.) However, this distribution does not fully describe the prior information about the setup; the very fact that the information is otherwise indifferent to the functional form of $f(x)$ has its own consequence which must not be overlooked. In general, it is important to take into account what is left unspecified by the prior information as carefully as what is specified, because the former automatically implies certain invariance properties related to the unspecified circumstances.

Now the change in the prior information produced by the change $x \rightarrow x+dx$ is represented by the statistical distance \cite{Fisher, Bhattacharyya, Wootters} between the corresponding probability distributions, namely,
\begin{equation}
ds^2= \sum_{i} (d \sqrt {P_i}\ )^2 = \frac{{f^\prime}^2}{4f(1-f)}\ dx^2. 
\end{equation}
This is nonzero, a manifestation of the fact that the information is not indifferent to the value of the path difference. However, because of the indifference of the prior information to $f(x)$, $|ds/dx|$ (which represents how the information is distributed with respect to $x$) is left (a priori) unspecified in this setup; in clear contrast with the situation in setup 1. In view of this unspecified circumstance, the most symmetric and least biased criterion would be to assign, a priori, the same value to $|ds/dx|$ for all values of $x$. Otherwise, clearly, $|ds/dx|$ could not have been completely unspecified; we must have had some kind of prior information about it. This `invariance property' for the metric is just an example of `equal a priori rule'; it avoids bias while agreeing with whatever information given. It is maximally noncomittal with regards to the unavailable information; any other criterion would introduce unconscious arbitrary assumptions not warranted by the information content of the setup. Therefore, the proper description of this information is provided by the criterion that 
\begin{equation}
\frac{|f^\prime|}{\sqrt{f(1-f)}}= k, 
\end{equation}
where $k$ is a (positive) constant related to the setup under consideration. This implies the solution,
\begin{equation}
f(x)=\frac{1}{2} (1 \pm \cos kx),
\end{equation}
so that
\begin{equation}
P_{1,2}= \frac{1}{2} (1 \pm \cos kx). \label{one}
\end{equation}
Now, on physical grounds, we know a priori that for a $x$-dependent probability distribution to emerge, a finite length scale is needed, in the abscence of which the experimental situation must remain scale-invariant. Thus, $k^{-1}$, is to be identified with some relevant property of the setup which, by preparation, has the same value in every repetition of the experiment, acting as a length scale to make $kx$ dimensionless. Its value is determined a priori from the preparation information. Given that no such provision has been made in the experimental preparation beforehand, we must set $k=0$, yielding, $P_{1,2}= 0,1$. On the other hand, given that it is the momentum, $p$, of the particle that is kept fixed in the experiment, we identify $k$ with $p$ (in units $\hbar=1$). Otherwise, i.e. when we do not know the manner of preparation, $k$ remains a free parameter to be determined experimentally; it will turn out to be $0$ or $p$, depending on the preparation of the experiment. In any case, the experimental distribution of the outcome will reflect this information as a part of the preparation information of the setup.
 
Result \ref{one} is the only probability distribution compatible with the information content of setup 2 and is, therefore, the distribution we expect to observe experimentally. This is why when the information is tampered with in anyway (as in delayed choice experiments), the outcome must change accordingly. Needless to emphasize, the notion of wave-particle duality, thus, becomes superfluous. The need to resort to wave nature (interference) not only does not arise to obtain the result but also such physical mechanisms, as explained, need not enter its interpretation at all. What represents the particle is an individual detector click and what represents the information content of the setup is the distribution of such clicks. In fact, as far as physical implications are concerned, this experiment is as trivial as setup 1.

It is not difficult to see that the above approach can be adopted to other experiments that have two potential final results at each repetition. For such experiments it is, thus, possible to obtain the standard quantum mechanical probabilities on the basis of their information content, rendering the superposition principle superfluous. 

The following comment may be in order. The above approach suffers from the same drawback that has always plagued the use of the ``equal a priori rule". Inspite of its large domain of application, it has the obvious difficulty that it is not invariant under a change of parameters and there seems to be no criterion for telling us which parametrization to use \cite{Jaynes, Jaynes2}. For exapmle, in our argument, if we were to parametrize the probabilities by $y=\sqrt{x}$ or $x^2$ instead of $x$, we would find that the probabilities vary sinusoidally with y, which is not the correct physical result. For the present, we propose to dodge this issue by regarding the approach as a naive form of the (presumably more fundamental) theory of optimal processing of information which would involve pondering into the nature of phase in terms of its role in the proper representation of information. Therefore, the value of the argument mainly lies in serving to stress the informational nature of quantum mechanics in the first place and to bring the underlying philosophy into focus.

\begin{acknowledgments}
I would like to acknowledge the hospitality of the Physics Department of Simon Fraser University where this work was carried out during my sabbatical leave. I would also like to thank Professor L. E. Ballentine for reading the manuscript.
\end{acknowledgments}

\end{document}